# Staircase-like metamagnetic transitions in phase-separated manganites: influence of thermal and mechanical treatments

V. Hardy, C. Yaicle, S. Hébert, A. Maignan, C. Martin, M. Hervieu, and B. Raveau.

Laboratoire CRISMAT, UMR 6508, ISMRA et Université de Caen, 6 Boulevard du Maréchal Juin, 14050 Caen, France.

## Abstract

Substitutions in the Mn-sublattice of antiferromagnetic, charge and orbitally ordered manganites was recently found to produce intriguing metamagnetic transitions, consisting of a succession of sharp magnetization steps separated by plateaus. The compounds exhibiting such features can be divided in two categories, depending on whether they are sensitive to thermal cycling effects or not. One compound of each category has been considered in the present study. The paper reports on the influence of two treatments: high-temperature annealing and grinding. It is shown that both of these treatments can drastically affect the phenomenon of magnetization steps. These results provide us with new information about the origin of these jumps in magnetization.

## Introduction

Mixed-valent perovskite manganites exhibit a great variety of fascinating properties, such as charge ordering (real-space ordering of the $Mn^{3+}$ and $Mn^{4+}$ species) and colossal magnetoresistance (collapse of the resistivity by several orders of magnitude under application of an external magnetic field) [1]. In lots of these materials, a phenomenon of electronic phase separation occurs, which influences considerably the physical properties [2-11].

Recently, peculiar metamagnetic transitions have been reported in Mn-site substituted manganites of general formula $Pr_{1-x}Ca_xMn_{1-y}M_yO_3$ (where M is a magnetic or non-magnetic cation, x is close to 0.5 and y is of the order of 0.05) [12-18]. After a zero-field cooling (ZFC), the M(H) curves at very low temperatures (typically below 5 K) exhibit a succession of steep jumps separated by plateaus. In large enough fields, the ferromagnetic state is reached. In these compounds, the ground state after ZFC essentially consists of an antiferromagnetic, charge and orbitally ordered (AFCOO) phase, that is noticeably weakened by the substitutions in the Mn sublattice. In most cases, this disordering effect also induces the development of ferromagnetic (F) domains that co-exist with the weak AFCOO state, leading to a phase-separated ground state. In such a competition, the application of a magnetic field favors the growth of the F phase at the expense of the AFCOO matrix. The peculiarity of the $Pr_{1-x}Ca_xMn_{1-y}M_yO_3$ compounds is the "jerky" character of their field-induced transitions at very low temperatures. In previous studies [15,18], we have discussed the likely role of the anisotropic strains that develop at the interfaces between the two phases due to their different unit cells. It was found that the appearance of magnetization jumps can be qualitatively explained by a competition between magnetic and elastic energies, in a process bearing some similarities with martensitic transitions in metallic alloys [18]. The present paper reports on two new kinds of experiments (high temperature annealing and modification of the microstructure by grinding) that were undertaken in order to test further the relevancy of such a martensitic scenario to the special staircase-like shape of M(H) curves in $Pr_{1-x}Ca_xMn_{1-y}M_yO_3$.

## Experimental details

This study has been carried out on two Mn-site substituted manganites: $Pr_{0.5}Ca_{0.5}Mn_{0.97}Ga_{0.03}O_3$ and $Pr_{0.6}Ca_{0.4}Mn_{0.96}Ga_{0.04}O_3$, hereafter denoted as [PrCa50]Ga3% and [PrCa40]Ga4%, respectively. The former compound is representative of the greatest part of these Mn-site substituted manganites, for which pronounced magneto-thermal cycling effects occur [13,18], whereas the latter compound is almost insensitive to such effects [15]. Ceramic samples of these compounds were prepared following standard solid state reaction described previously [12].

Magnetic measurements were performed by means of commercial magnetometers using extraction methods (MPMS and PPMS from Quantum Design). In all cases, the M(H) curves were recorded after a zero-field cooling from 300 K. For both compounds, the transition temperatures related to spin, charge or orbital ordering are larger than 300 K [13,15]. Hysteresis loops were collected by using a field spacing equal to 0.25 T and a waiting time of 1 minute after each field installation.

Annealings were carried out in air at various temperatures ($T_A$=400 °C; 600 °C; 800 °C; and 1000 °C). In all cases, the samples were introduced into a furnace previously stabilized at $T_A$, and they were maintained at this temperature 10 hours. Then, the samples were either directly taken out of the furnace (referred to as the Quench mode), or they were slowly cooled down to room temperature in 48 hours (referred to as the Slow mode).

Ceramic bars were manually ground in an agate mortar. The distribution of the grain sizes was analyzed by Scanning Electron Microscopy (Philips XL30) on a fractured face of a ceramic bar, while it was investigated by granulometry (Mastersizer, Malvern Instruments) for the powder obtained after the grinding. An average grain size close to 20 µm was found in both cases. It is quite unlikely, however, that the grinding just dissociated the grains of the ceramic without breaking at least some of them. It turns out that the quantitative comparison of the measurements obtained before and after the grinding is delicate since they are derived from very different techniques. Actually, it must be expected that the average grain size is smaller after the grinding than before, but our measurements revealed that the difference remains moderate.

## Results

Figure 1 shows M(H) curves recorded at 2.5 K in [PrCa50]Ga3% and [PrCa40]Ga4%. These curves are characteristic of the staircase-like field-induced transitions occurring in these compounds at low temperatures. In both compounds, the field-increasing branches of the loops exhibit abrupt jumps separated by plateaus, and the magnetization in the highest field reaches a value close to that expected for a full polarization of the Mn-spins (3.38 and 3.44 $\mu_B$/f.u. in [PrCa50]Ga3% and [PrCa40]Ga4%, respectively). One can also observe a rounding of the field-increasing branches in the low-field regime, a feature related to the ferromagnetic domains that are present in the ground state of these phase-separated systems. Note that such a ferromagnetic component is also found in [PrCa50]Ga3%, though it is much smaller than in [PrCa40]Ga4%. The volume fraction of ferromagnetic phase in zero-field ($f_0$) can be derived from the non-linear part of the virgin magnetization curve under low fields [13]. The $f_0$ values in [PrCa50]Ga3% and [PrCa40]Ga4% are found to be close to 1% and 25%, respectively.

Each panel of Fig. 1 displays the first and the last curve of a series of four ZFC loops that were successively recorded on a virgin sample. The case of [PrCa50]Ga3% illustrates the spectacular cycling effects that are encountered in most cases. These cycling or training effects consist of (i) a decrease in the slope of M(H) below the first step field; (ii) changes in the location of the step fields and in the magnetization values of the plateaus; (iii) a decrease in the value of magnetization in the highest field available. The situation is very different in [PrCa40]Ga4% for which no significant cycling effects can be detected.

In [PrCa50]Ga3%, annealing experiments have been carried out in order to try to recover the properties of the virgin state. Firstly, several samples were "trained" by recording four successive loops at 2.5 K. Secondly, they were annealed at different temperatures $T_A$ (400 °C; 600 °C; 800 °C; 1000 °C). Since the last stage of the synthesis of [PrCa50]Ga3% is a plateau at 800 °C followed by a quench to room temperature, we used this Quench mode in the first series of experiments. Finally, a new M(H) curve was recorded on each of these annealed samples. For $T_A \geq 600$ °C, the new curves obtained after annealing are very close to that recorded in the virgin state (first loop). The jumps occur at the same fields as in the first loop, while the zero-field ferromagnetic fraction ($f_0$) and the magnetization in 9 T ($M_{max}$) recover more than 90% of their initial values. Although the differences between 600, 800 and 1000 °C are very small, the best results are found for $T_A$ =800 °C regarding the $f_0$ and $M_{max}$ values. The superimposition onto the virgin curve can still be slightly improved by using the Slow cooling mode. Figure 2(a) shows the first and fourth loop recorded on a virgin sample of [PrCa50]Ga3%, along with the curve obtained after an annealing at 800 °C followed by a Slow cooling. Figure 2 (b) shows that the annealing at $T_A$ =400 °C has practically no "regenerating" effect.

Figure 3(a) shows M(H) curves at 5 K for [PrCa50]Ga3%, that were recorded on the same sample before and after it had been ground. The steps that were clearly visible on the curve of the ceramic bar have totally disappeared after the grinding. One can also observe that the zero-field ferromagnetic fraction ($f_0$) is decreased by this treatment. In this experiment, it can be noticed that the measurement of the powder is performed on a sample that has already experienced a magneto-thermal cycling [19]. To clearly separate the grinding effect from a possible influence of the training effect discussed above, another virgin ceramic bar was directly powdered and then measured for the first time. The M(H) curve obtained in this way is well superimposed on the previous one measured on the powder [see the dashed line in Fig. 3(a)]. This ensures that the spectacular change of shape shown in Fig. 3(a) is really related to the microstructure.

Similar effects have been found at lower temperature, except that a careful inspection of the curves at 2.5 K reveals the persistence of small step-like features in the ground sample [Fig. 3(b)]. This result is of importance since it indicates that the grinding does not merely suppress the steps, but it rather seems to limit their appearance to a lower temperature range, and their amplitude to smaller values.

Figure 4 shows M(H) curves at 2.5 K for [PrCa40]Ga4%, that were recorded on the same sample before and after the grinding. Similar results were found at 5 K. In [PrCa40]Ga4%, there are still large magnetization jumps after the grinding and the magnetization still reaches the complete saturation in highest fields. It must be emphasized, however, that the location of the step fields and the values of magnetization at the plateaus are affected by the grinding. Moreover, the grinding produces a pronounced decrease in the initial ferromagnetic fraction $f_0$, as exhibited by the marked change in shape of the field-increasing branch in low field.

## Discussion

Most of the Mn-site substituted manganites showing staircase-like metamagnetic transitions are subject to pronounced training effects, as illustrated in Fig. 1(a) with the case of [PrCa50]Ga3%. Let us stress again that, in these experiments, the samples are systematically zero-field cooled from 300 K before each M(H) curve, a temperature that is well above all the magneto-electronic transitions occurring in these materials. As many other Mn-site substituted manganites [20], the $Pr_{1-x}Ca_xMn_{1-y}M_yO_3$ compounds are phase-separated systems in which can co-exist ferromagnetic (F) and antiferromagnetic, charge and orbitally ordered (AFCOO) domains [21]. Their ground state is generally complex, and it can contain several antiferromagnetic phases. In the case of [PrCa50]Ga3% for instance, a recent neutron diffraction study [22] has shown the presence of two antiferromagnetic phases, one of CE type and the other of pseudo-CE type. Both of them exhibit orbital ordering, and though the existence of charge ordering in the manganites is currently debated, we still refer to them as AFCOO phases hereafter, for the sake of simplicity.

Each time a M(H) curve is recorded, two processes successively affect the balance between the ferromagnetic (F) and the antiferromagnetic (AFCOO) phases: firstly, a certain volume fraction of ferromagnetic domains ($f_0$) grows within the AFCOO matrix along the zero-field cooling; secondly, the application of magnetic field at low T tends to switch the sample into the F state. A fact of crucial importance is that the structural unit cells of the F and AFCOO phases are different. Let us evaluate the degree of distortion of the unit cell via the parameter $D = (100/3) \sum_{i=1,2,3} \left( |a_i - \langle a \rangle| / \langle a \rangle \right)$, where $a_i$ refers to the three cell parameters, and $\langle a \rangle$ is the average value [23]. For instance, in the case of [PrCa50]Ga3% at 2.5 K, one has D ~ 0.22 and D ~ 0.76 for the pseudo-CE and CE phases (zero-field data), respectively, whereas the F phase which grows under field application (5.5 T data) is significantly less distorted, D ~ 0.12. Owing to such pronounced differences between the units cells of the AFCOO and F phases, the growth of the F phase at the expense of the AFCOO phases must produce structural distortions, and large strains must develop at the interfaces between the two kinds of domains. These strains may create structural defects, which are then susceptible to affect the establishment of the phase-separation in subsequent measurements [18]. Training effects associated with the defects creation are well known in standard martensitic transitions [24]. In our case, one may speculate that the decrease of $f_0$ seen in Fig. 1(a) and Fig. 2 results from the influence of such defects, that impede the development of the ferromagnetic fraction in zero-field. In this scenario, the evolution of the step pattern in [PrCa50]Ga3% is ascribed to the decrease in $f_0$, a feature, which is itself, caused by the defects created along previous measurements.

We have seen that simple annealings at temperatures equal to or above 600 °C can restore the initial properties of the sample. One of the main effects that are expected for this type of annealing in oxides is the removal of structural defects. Therefore, the spectacular influence of these annealings is found to be well consistent with the scenario mentioned above. In addition, the absence of effect for $T_A = 400$ °C indicates that atomic displacements are required to restore the sample in its virgin state, as it is expected if one needs to remove structural defects. At the present time, the precise nature of these defects remain to be explored. Let us just note that a preliminary electron microscopy study has been carried out in two steps. First for the two pristine samples, it was observed at room temperature that the introduction of foreign atoms in the manganese sites does not generate specific extended defects. At 92K, the existence of complex CO phenomena, with incommensurate modulated structures, was evidenced. Second, no significant variation of the microstructural state of the crystals has been detected at room temperature in the sample [PrCa50]Ga3% which suffered 6 cycles at 5K.

We have shown that the peculiar shape of the M(H) curves at low T in $Pr_{1-x}Ca_xMn_{1-y}M_yO_3$ compounds can be profoundly affected by a simple grinding of ceramic samples. Two main effects are expected from such a mechanical treatment: (i) a disconnection of the grains of the ceramic; (ii) a reduction of their average size. To check the absence of any influence of field-induced alignment of the grains in the powder, a M(H) curve was recorded in a sample made of compacted powder. The results were the same as those shown in Fig. 3 for the non-compacted powder. Therefore, the spectacular changes displayed in Fig. 3 should be connected with the decrease in the average grain size, though our measurements suggested a rather small modification. It has been already shown that moderate variations in the grain size can modify drastically some properties of phase-separated manganites, as for instance the insulator to metal transition under zero field [25]. As emphasized by Podzorov *et al.* [25], the microstructure can be a determinant parameter since the martensitic strains related to phase separation are more difficult to accommodate as the grain size decreases. In our case, this phenomenon can account for the fact that the zero-field ferromagnetic fraction (resulting from the growth of F domains within the AFCOO matrix) is strongly reduced after the grinding. Note that this effect also exists in [PrCa50]Ga3%, although it is less visible because of the small $f_0$ value of the ceramic.

In addition to the decrease in $f_0$, the grinding affects the development of the magnetization steps at low temperatures. Comparison of Fig. 3 and Fig. 4 shows that this effect is more pronounced in the case of [PrCa50]Ga3%. It is worth noticing that [PrCa50]Ga3% also exhibits sizeable training effects, in contrast to the case of [PrCa40]Ga4%. When one compares the properties of virgin ceramic samples, one of the most obvious difference between these two compounds concerns the zero-field ferromagnetic fraction that is much smaller in [PrCa50]Ga3% than in [PrCa40]Ga4%. Within the framework of the competition existing between the AFCOO and the F phases, this weaker ferromagnetic tendency of [PrCa50]Ga3% may be the reason why this compound is more sensitive than [PrCa40]Ga4% to all perturbations tending to impede the establishment of ferromagnetic domains (like cycling or grinding).

Beyond the existence or not of magnetization steps in [PrCa50]Ga3% (i.e. at 2.5 or 5 K), the ferromagnetic fraction in the ground sample only increases slowly with the magnetic field (see Fig. 3). At 5 K, the thermal energy may be high enough to allow the accommodatation of such smooth variations in ferromagnetic fraction, whereas small steps in magnetization can emerge at lower temperatures (2.5 K). In [PrCa40]Ga4%, the initial ferromagnetic fraction is considerably reduced by the grinding but it keeps a sizeable value. The persistence of a large ferromagnetic fraction under zero-field in powdered [PrCa40]Ga4% demonstrates that the magnetic energy favouring the development of ferromagnetic domains remains large enough to overcome the elastic energy associated with the martensitic strains (despite the enhancement of this last term due to the grinding). Thus the situation in the ceramic and in the powder remain quite close to each other in the case of [PrCa40]Ga4%. Accordingly, large steps are also observed in the ground sample under application of field at low temperature. It must be emphasized, however, that the locations of the step fields are different than in the ceramic sample. This may be just due to the different accommodation ability of the martensitic strains in the powder compared to the ceramic.

## Conclusion

This study shows that the staircase-like metamagnetic transitions in $Pr_{1-x}Ca_xMn_{1-y}M_yO_3$ compounds can be strongly affected by both high temperature annealing and grinding. In samples where the steps progressively disappear with cycling [successive M(H) loops recorded after ZFC], it was shown that the sample recovers its initial properties by annealing at temperatures close to 800 °C. Note that in this range of temperature, grain boundaries are regenerated and that major structural mechanims can be activated. Importantly, it was observed that a simple grinding of ceramic samples severely alters the pattern of steps. In the most sensitive compounds (illustrated here by [PrCa50]Ga3%), the grinding leads to a disappearance of the magnetization jumps at the highest temperatures, while it considerably reduces their amplitude at the lowest temperatures. In the least sensitive samples (illustrated here by [PrCa40]Ga4%), large magnetization jumps are still present after the grinding, but the locations and the heights of the steps are changed.

Both the training and grinding effects can hardly be explained if the steps are only related to some intrinsic phenomena, such as field-induced transitions in the magneto-electronic ground state. Actually, the spectacular effect of a simple grinding clearly demonstrates that the step fields cannot be regarded as true, intrinsic critical fields. Furthermore, one can consider that both series of results (annealing and grinding effects) point to a crucial role of the microstructure on the development of the magnetization steps.

On the one hand, it turns out that this new series of results is qualitatively consistent with the martensitic scenario [18] that has been proposed to be at the origin of the magnetization steps in $Pr_{1-x}Ca_xMn_{1-y}M_yO_3$ compounds. On the other hand, it must be emphasized that the nature of the defects that are invoked has not yet been elucidated, and the study of the effects related to the grain size and the grain boundaries would need a better control of the microstructural features. It is clear that further experiments will have to be carried out on each of these issues.

## Acknowledgements

The authors thank L. Hervé for SEM investigations, and J. Lecourt for granulometry measurements.

Figure Captions

Fig. 1: First (circles) and last (diamonds) curve of a series of four ZFC M(H) curves at 2.5 K that were successively recorded on the same sample. Upper panel shows the case of $Pr_{0.5}Ca_{0.5}Mn_{0.97}Ga_{0.03}O_3$. Lower panel shows the case of $Pr_{0.6}Ca_{0.4}Mn_{0.96}Ga_{0.04}O_3$. The arrows denote the field-increasing and the field-decreasing branches.

Fig. 2: Effects of two annealing temperatures on the cycling effect in $Pr_{0.5}Ca_{0.5}Mn_{0.97}Ga_{0.03}O_3$: (a) 800 °C; (b) 400 °C. For the sake of clarity, only the field-increasing branches of M(H) are shown: first curve (open circles) and fourth curve (open diamonds) on the virgin sample; curve after the annealing (closed circles).

Fig. 3: Grinding effect in $Pr_{0.5}Ca_{0.5}Mn_{0.97}Ga_{0.03}O_3$: (a) M(H) at 5 K before (open circles) and after (closed circles) the grinding; (b) same curves at 2.5 K. The dashed line in (a) is the M(H) curve recorded in another sample which has been ground without prior measurements. The arrows in (b) emphasise the occurrence of small magnetization steps at 2.5 K.

Fig. 4: M(H) curves at 2.5 K in a $Pr_{0.6}Ca_{0.4}Mn_{0.96}Ga_{0.04}O_3$ sample, before (open circles) and after (closed circles) the grinding.

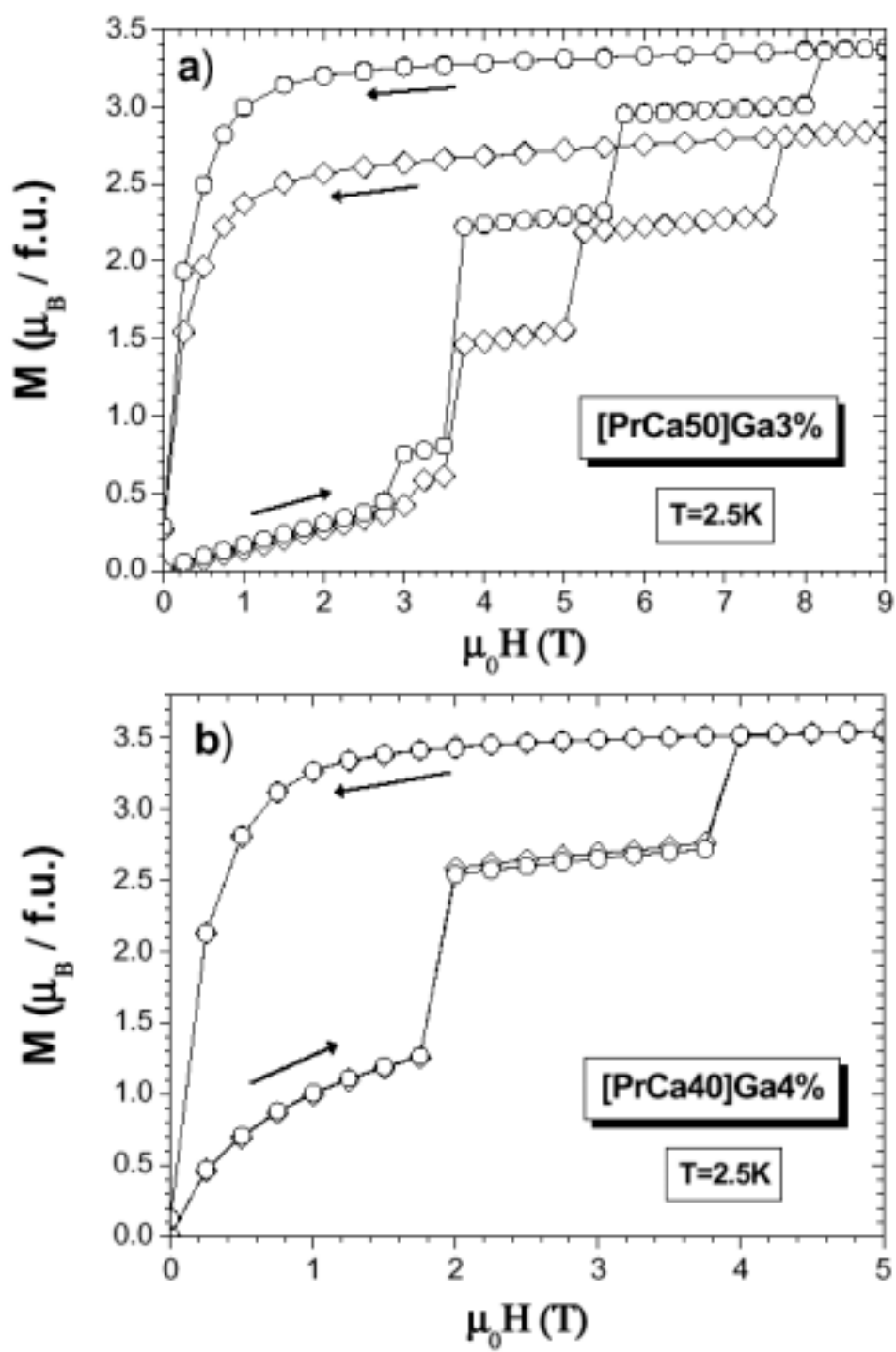

Figure 1

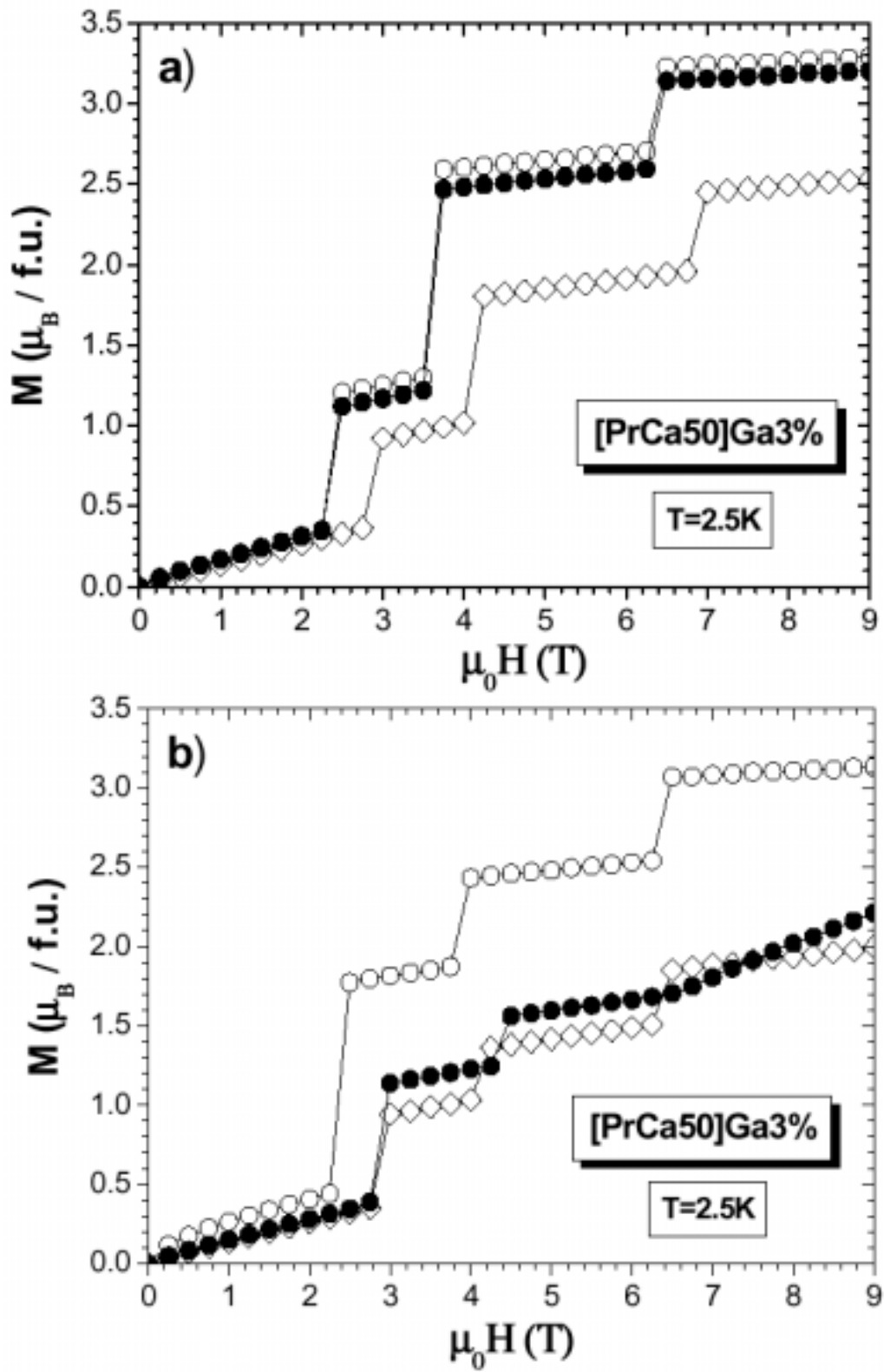

Figure 2

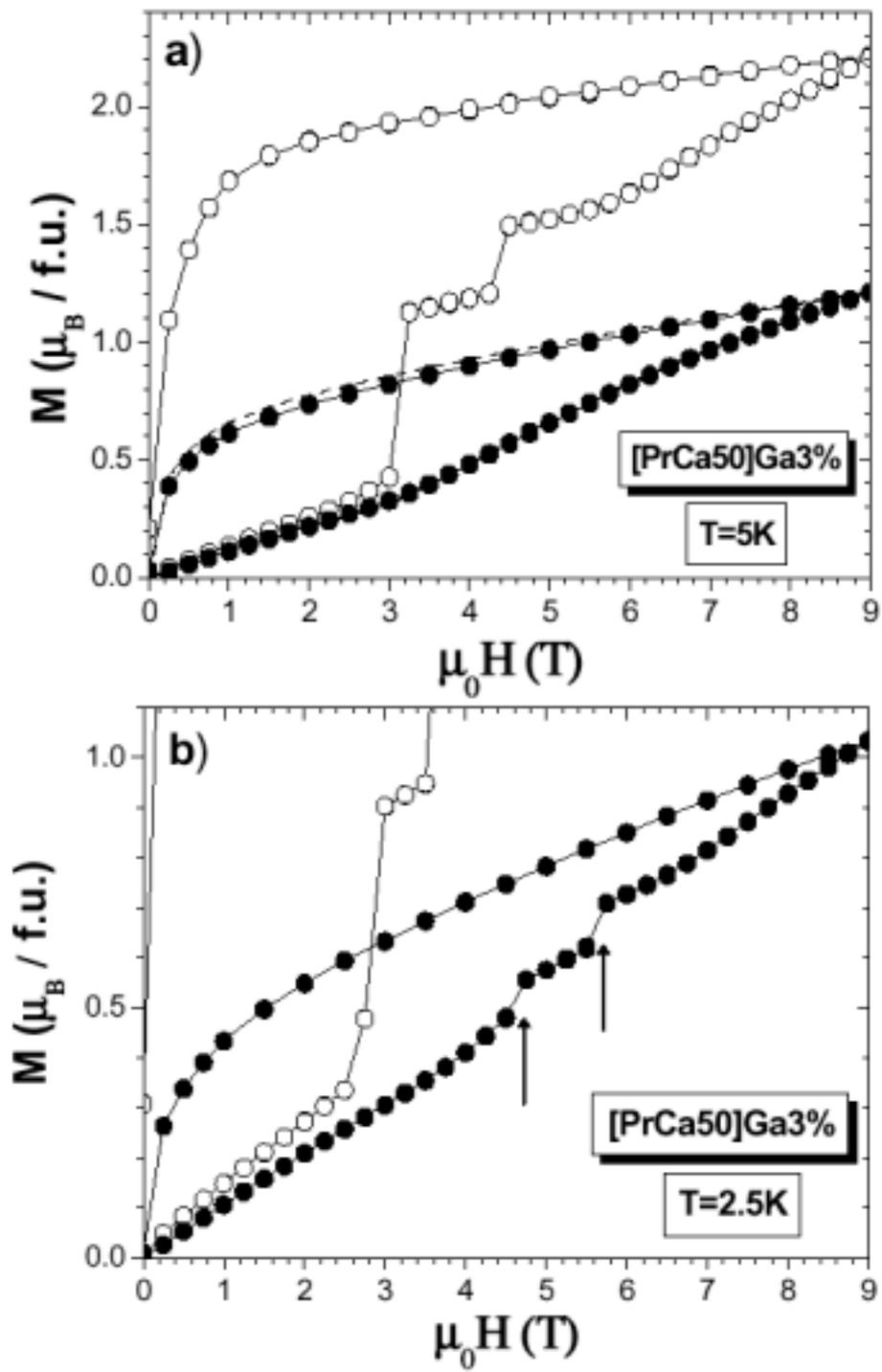

Figure 3

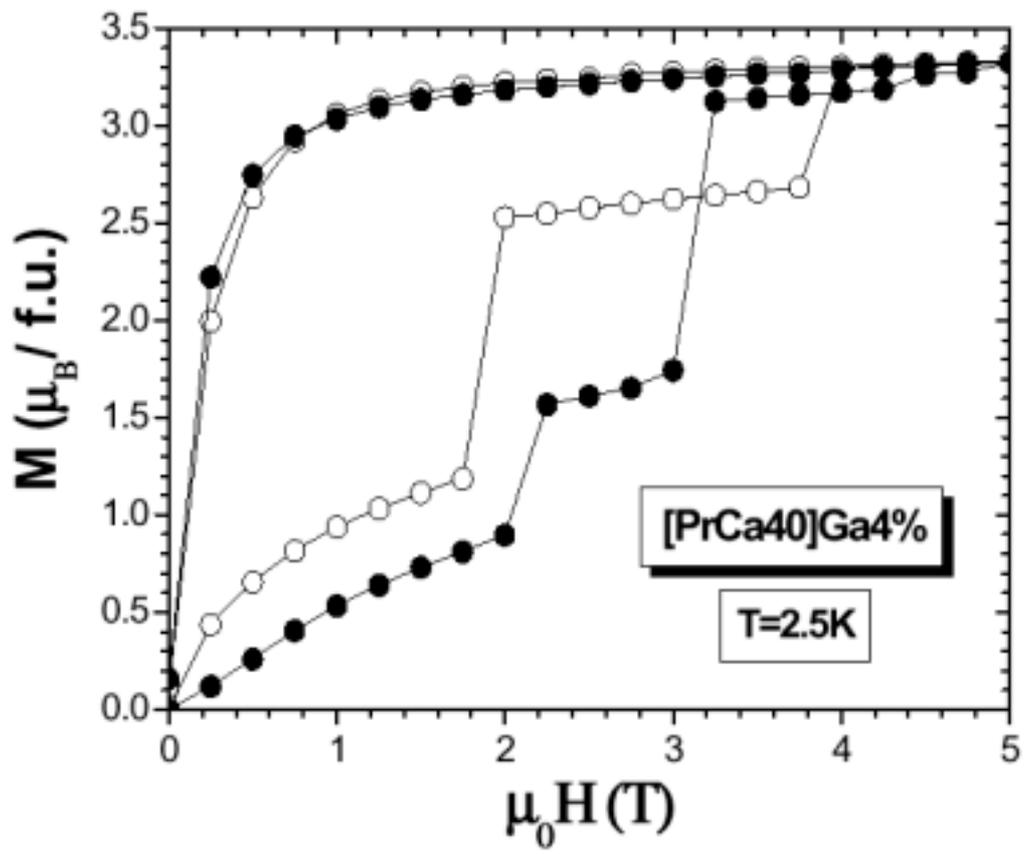

Figure4